\documentclass[main]{aa}  
\usepackage{graphicx}
\usepackage{txfonts}
\usepackage{amsmath}
\usepackage[utf8]{inputenc}
\usepackage{kotex}
\usepackage{xcolor}
\usepackage{verbatim}

\bibpunct{(}{)}{;}{a}{}{,} 

\def\apjl{{ApJL}}
\def\apjs{{ApJS}}

\def\hal{H$\alpha$}
\def\hb{H$\beta$}

\def\redd{$\lambda_{\rm Edd}$}
\def\lax{{$\mathrel{\hbox{\rlap{\hbox{\lower4pt\hbox{$\sim$}}}\hbox{$<$}}}$}}
\def\gax{{$\mathrel{\hbox{\rlap{\hbox{\lower4pt\hbox{$\sim$}}}\hbox{$>$}}}$}}
\def\simlt{\lower.5ex\hbox{$\; \buildrel < \over \sim \;$}}
\def\simgt{\lower.5ex\hbox{$\; \buildrel > \over \sim \;$}}

\def\cm2{cm$^{-2}$}

\def\feii{\ion{Fe}{II}}
\def\mgii{\ion{Mg}{II}}
\def\civ{\ion{C}{IV}}

\def\lbol{$L_{{\rm bol}}$}
\def\ledd{$L_{{\rm Edd}}$}

\def\l5100{$L_{5100}$}
\def\ll5100{$\log L_{\rm 5100}$}

\def\llbol{$\log\ (L_{\rm bol}/{\rm erg\ s^{-1}})$}

\def\lredd{$\log \lambda_{\rm Edd}$}
\def\-->{$\rightarrow$}
\def\sf200{${\rm SF}_{\rm 200 days}$}

\begin{document}

\title{Temperature profiles of accretion disks in luminous active galactic nuclei derived from ultraviolet spectroscopic variability
}

\author{Suyeon Son\inst{1,2} \and Minjin Kim\inst{1} \and Luis C. Ho\inst{2,3}}

\institute{Department of Astronomy and Atmospheric Sciences,
Kyungpook National University, Daegu 41566, Korea\\
\email{mkim.astro@gmail.com}
\and
Kavli Institute for Astronomy and Astrophysics, Peking University, Beijing 100871, China
\and
Department of Astronomy, School of Physics, Peking University, Beijing 100871, China
}

\date{Received}

\abstract{
The characteristic timescale ($\tau$) of the continuum variability of the accretion disk in active galactic nuclei (AGNs) is known to be related to the thermal timescale, which is predicted to scale with AGN luminosity ($L$) and the rest-frame wavelength ($\lambda_{\rm RF}$) as $t_{\rm th} \propto L^{0.5} \lambda_{\rm RF}^2$ in the standard disk model. Using multi-epoch spectroscopic data from the Sloan Digital Sky Survey Reverberation Mapping project, we constructed ultraviolet ensemble structure functions of luminous AGNs as a function of their luminosity and wavelength. Assuming that AGNs exhibit a single universal structure function when $\Delta t$ is normalized by $\tau$, wherein $\tau \propto L^{a} \lambda_{\rm RF}^{b}$, we find $a=0.50\pm0.03$ and $b=1.42\pm0.09$. While the value of $a$ aligns with the prediction from the standard disk model, $b$ is significantly smaller than expected, suggesting that the radial temperature (color) profile of the accretion disk is significantly steeper (shallower) than the standard disk model. Notably, this discrepancy with theory has been observed in previous studies based on spectroscopic reverberation mapping and gravitational microlensing. Although no current model of accretion disks fully matches our results, our findings provide valuable constraints for testing future physical models.
}


\titlerunning{Temperature Profiles of AGN Accretion Disks}
\authorrunning{Son et al.}

\maketitle

\section{Introduction} 
While supermassive black holes (BHs) are ubiquitous in massive galaxies, only a minority of them accrete sufficiently to shine as active galactic nuclei (AGNs). AGNs emit strong radiation over a wide range of wavelengths, from the radio to X-rays. In general, the emission from AGNs exhibits substantial variability with a timescale of days to years across all energies. In addition, variability originating from different substructures is found to be correlated \citep{matthews_1963,wagner_1995,ulrich_1997}. Under the assumption that the intrinsic variation in the accretion disk is echoed by extended structures such as the broad-line region (BLR) and the dusty torus with a time delay corresponding to the light crossing time, the radial size of such structures can be estimated by measuring time lags. Reverberation mapping (RM; \citealt{bahcall_1972,blandford_1982}), first applied to probe the size of the BLR using the time delay between the variation in the accretion disk continuum and the broad emission lines \cite[e.g.,][]{peterson_1994, obrien_1998}, led to the development of methods for estimating the masses of BHs in AGNs \cite[e.g.,][]{gebhardt_2000, peterson_2004, greene_2005, vestergaard_2006, ho_2015,wang_2024}.

The RM method also provides a powerful probe of the size of the accretion disk \citep{krolik_1991,wanders_1997}. Simplifying the structure of an accretion disk into a series of annuli with a negative temperature gradient (i.e., lower temperatures at larger radii), the time lag between the accretion disk continuum at different wavelengths, originating from different temperatures, can serve as a proxy for the disk size. Intriguingly, disk sizes estimated from continuum RM are systematically larger (e.g., \citealt{fausnaugh_2016}; but see \citealt{homayouni_2019}) than predicted from the standard disk model \citep{shakura_1973, frank_2002}. A similar trend has also been identified through gravitational microlensing studies (e.g., \citealt{morgan_2010}). Various explanations have surfaced to explain this ``disk size problem,'' most of which are based on a modification of the geometry or emitting source: an accretion disk with a disk wind \citep{sun_2019}, a rimmed or rippled accretion disk \citep{starkey_2023}, an X-ray source with an increased height \citep{kammoun_2019}, a disk reprocessing in the X-rays or far-ultraviolet \citep{gardner_2017}, or a significant contribution from the recombination continuum of the BLR (e.g., \citealt{mchardy_2018}).

The radial profile of the effective temperature, defined as $T \propto R^{-1/p}$, can be used to test the validity of accretion disk models, as different models predict different temperature profiles. From an observational point of view, under the assumption that the radiation follows a blackbody spectrum ($\lambda_{\rm max} \propto T^{-1}$), it is more common to measure the more observationally tractable radial color profile, $R \propto \lambda_{\rm RF}^p$, where $\lambda_{\rm RF}$ is the rest-frame wavelength. For example, while $p=4/3$ is expected from the standard disk model, a larger value of $p$ is anticipated from the disk wind model \citep{li_2019,sun_2019}, and the opposite holds for the magnetorotational instability model \citep{agol_2000}. Continuum RM studies find that the observed time lags for various wavelengths are well fit with $p \approx 4/3$, as expected from the standard disk model (\citealt{mchardy_2014,homayouni_2019}). However, both steeper ($p<4/3$; \citealt{fausnaugh_2016,starkey_2017}) and shallower ($p>4/3$; \citealt{jiang_2017}) temperature profiles have been detected. The results from microlensing studies, which report a wide range of values for $p$, are more debated. While many claim that $p$ is consistent with $4/3$ (e.g., \citealt{eigenbrod_2008,poindexter_2008,rivera_2023}), values of $p$ lower than $4/3$ have been derived by \cite{floyd_2009}, \cite{blackburne_2011}, \cite{jimenez_2014}, and \cite{munoz_2016}, and values of $p$ higher than $4/3$ have been derived by \cite{bate_2018} and \cite{cornachione_2020}. The origin of these discrepant results is unclear. Certainly, small-number statistics come into play, and it is difficult to judge the impact of sample selection and experimental methodology. 
It is vital to investigate the radial color profile for a large sample of AGNs using an independent method.

As a complementary approach, color profiles can be quantified using the characteristics of ultraviolet-optical variability arising from the accretion disk. This intrinsic variability can be characterized using the power spectral density, defined as the variability power as a function of the frequency ($\nu$). One can also compute the structure function (SF), defined as the variability amplitude as a function of the time lag ($\Delta t$). The power spectral density and SF of the accretion disk continuum are well fit with a broken power law in the sense that the variability amplitude increases with increasing timescale and flattens above a certain characteristic timescale ($\tau$; \citealt{kelly_2009,macleod_2010,guo_2017}). Studies have shown that $\tau$ corresponds to the thermal timescale\footnote{The thermal timescale is the approximate time it takes to establish thermal equilibrium.}, $t_{\rm th}$ (\citealt{kelly_2009,sun_2015,burke_2021,arevalo_2024}; but see \citealt{macleod_2010}). Although most studies reached a similar conclusion, the relations they found between $\tau$ and AGN properties vary, possibly due to systematic uncertainties in the measurements of $\tau$ caused by imperfect light curves with a sparse cadence and a relatively short baseline \citep{emmanoulopoulos_2010,kozlowski_2017}.

Alternatively, the dependence of $\tau$ on the AGN luminosity and wavelength can be investigated by assuming that all the SFs or power spectral densities follow a single universal function if $\Delta t$ or $\nu$ is converted to the units of the characteristic timescale. Applying this method to the broadband light curves of thousands of AGNs, \citet{tang_2023} demonstrated that $\tau \propto L^{0.539\pm0.004}\lambda_{\rm RF,broad}^{2.418\pm0.023}$, where $\lambda_{\rm RF}$ is the effective wavelength of the broadband filter in the rest-frame. Although the slopes of the luminosity and wavelength factor are marginally steeper than expected from the standard disk model, this finding is broadly consistent with the predictions of the standard disk model on the thermal timescale (i.e., $\tau \propto L^{0.5}\lambda_{\rm RF}^{2}$; but see \citealt{arevalo_2024,petrecca_2024}). It should be noted that, in the standard disk model, the thermal timescale is expected to correlate with the AGN luminosity and wavelength as $\tau \propto L^{0.5} \lambda_{\rm RF}^2$ \citep{shakura_1973,frank_2002}. This discrepancy is possibly due to several shortcomings of broadband photometry, including (1) contamination from broad emission lines and iron multiples \cite[e.g.,][]{boroson_1992}, (2) the broad range of the effective wavelength in the rest-frame, which depends on the redshift of the targets \cite[e.g.,][]{macleod_2010}, and (3) time variation of the shape of the continuum \cite[e.g.,][]{guo_2016}, which can introduce bias when converting broadband photometry to continuum flux. Despite these limitations, broadband monitoring data have been widely used, for they can be obtained for large samples more efficiently than the more resource-intensive spectroscopic monitoring, which is available only for more limited samples.

To properly investigate the dependence of $\tau$ on the rest-frame wavelength, with the aim of constraining the temperature profile of the accretion disk, it is preferable to measure the continuum variability using spectroscopic data, which can more effectively  mitigate contaminants such as strong emission lines. The Sloan Digital Sky Survey Reverberation Mapping (SDSS-RM) project \citep{shen_2015a, shen_2024}, which has conducted intensive spectroscopic monitoring of a relatively large sample of quasars, provides an opportunity to carry out this experiment. With this dataset, we examined the dependence of the characteristic timescale of continuum variability on AGN luminosity and wavelength using spectroscopically determined continuum light curves. 

\begin{figure}[t]
\centering
\includegraphics[width=0.45\textwidth, page=1]{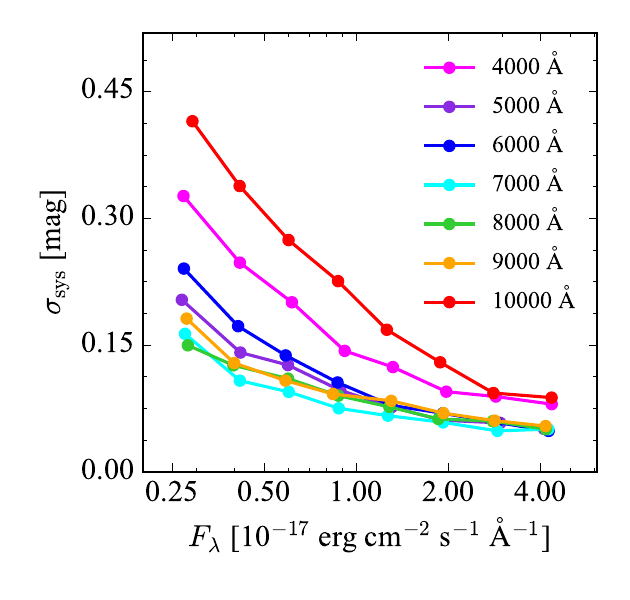}
\caption{Systematic uncertainty in magnitude due to the uncertainty in flux calibration ($\sigma_{\rm sys}$) as a function of the observed wavelength and flux density. The systematic uncertainties increase dramatically at the blue and red ends of the spectra ($\lambda_{\rm obs} < 5000\, {\rm \AA}$ and $\lambda_{\rm obs} > 9000\, {\rm \AA}$).
}
\end{figure}

\section{Sample and data}
The continuum variability of AGNs typically is studied using broadband photometry (e.g., \citealt{vandenberk_2004,macleod_2010,caplar_2017,tang_2023,arevalo_2024}). A shortcoming of this conventional method is that broadband photometry can easily be contaminated by strong emissions from the BLR, including broad emission lines, the myriad blends from iron multiplets, and the Balmer continuum. By comparison, spectroscopic data offer the advantage that specific wavelength windows can be chosen to reject strong contaminants. This is the strategy we adopted in this study. We utilized the spectroscopic monitoring data from the SDSS-RM project\footnote{\url{https://ariel.astro.illinois.edu/sdssrm/final_result/}} \citep{shen_2015a,shen_2024} to construct the ultraviolet continuum SF. The SDSS-RM project monitored a flux-limited sample ($i_{\rm PSF}<21.7$ mag) of 849 quasars at $0.1<z<4.5$ using a 2.5~m SDSS telescope with the Baryon Oscillation Spectroscopic Survey (BOSS) spectrograph, covering the wavelength range $3650-10400\,\rm{\AA}$ at a spectral resolution of $\sim 2000$. Light curves from this project span $\sim7$ years with an average cadence of $\sim4$, 12, and 30 days for the 2014, 2015$-$2017, and 2018$-$2020 observing seasons, respectively. Because accurate measurement of the continuum fluxes and their uncertainties is crucial to estimating the SF robustly, we computed those values directly from the spectra instead of extracting them from the best-fit continuum, which can lead to systematic bias.

\subsection{Measurements of flux and its uncertainty}

To measure the continuum variability using spectroscopic data, we defined continuum windows at various wavelengths and estimated the flux densities from the single-epoch spectra within these regions.
The uncertainties of the flux density are given in each spectral pixel for the single-epoch spectra. However, additional uncertainties stemming from imperfect flux calibration \cite[e.g.,][]{margala_2016,milakovic_2021} should be taken into account. We estimated the final uncertainty of the continuum flux ($\sigma_{\rm final}$) as the square root of the quadratic sum of these two sources of uncertainty.

The uncertainty measured from the continuum window with a width of 40 ${\rm\AA}$ was calculated as $\sigma_{\rm flux}/\sqrt{N}$, where $\sigma_{\rm flux}$ is half the 16th to 84th percentile range of the flux density, and $N$ is the number of data points within the continuum window. We used the median of the flux density within the continuum window as the flux at the continuum. We statistically estimated the systematic uncertainty due to the unstable flux calibration using the multi-epoch spectra of our sample. If we assume that our sample does not vary over a period of 2 days (i.e., ${\rm SF_{2\,days} \approx 0}$), the observed flux variation within this time interval is caused solely by the systematics in the flux calibration. This systematic uncertainty, in units of magnitude, can be expressed as $\sigma_{\rm sys}=\sqrt{(\sigma_{\rm \Delta mag}/\sqrt{2})^2-\sigma_{\rm mag}^2}$, where $\sigma_{\rm \Delta mag}$ is half the 16th to 84th percentile range of the differences between magnitude pairs observed within 2 days of one another\footnote{Note that the pair is selected in the observed frame. Therefore, in the rest-frame, the $\Delta t$ between the pair is on average $\sim0.76\pm0.21$ days.}, and $\sigma_{\rm mag}$ is the uncertainty of the continuum flux in the single-epoch spectrum in units of magnitude (Fig.~1). This value cannot be estimated for an individual object because there are typically fewer than seven spectroscopic pairs observed within $2$ days for a single target, which is not sufficient to estimate the uncertainty. Instead, we derived the systematic uncertainties using the entire sample as a function of the observed wavelength ($\lambda_{\rm obs}$) and flux density ($F_\lambda$). We find that it is sensitive to the observed wavelength and the brightness of the target (Fig.~1). The systematic uncertainty of the single spectra at a specific wavelength was determined via interpolation.

Intriguingly, $\sigma_{\rm sys}$ is substantial at the spectral edges (e.g., $\lambda_{\rm obs} < 5000\, {\rm \AA}$ and $\lambda_{\rm obs} > 9000 {\rm \AA}$), indicating that the relative flux calibration error and imperfect sky subtraction may be severe in the original data. Motivated by this result, we opted not to use the spectral data at $\lambda_{\rm obs} < 5000\, {\rm \AA}$ or $\lambda_{\rm obs} > 9000 {\rm \AA}$ in the observed frame. Additionally, we did not use objects with $\sigma_{\rm final} \geq 0.11$\,mag, for which the noise dominates the variability, because we cannot measure the error exactly for each object, we can only measure it statistically. Finally, to investigate the variability solely from the accretion disk, we attempted to avoid the spectral regions that can be highly contaminated by broad emission lines, the pseudo-continuum originating from \feii\ multiplets \citep{shen_2011}, and the stellar continuum of the host galaxy. Consequently, the continuum fluxes were measured as median values within 40-${\rm\AA}$-wide continuum windows centered at 1367, 1479, 1746, 2230, 3050, 3500, and 3800 ${\rm \AA}$ in the rest-frame. We note that the continuum fluxes at 3050 and 3500 \AA\ may be influenced by the Balmer continuum \cite[e.g.,][]{cackett_2018, korista_2019}. This phenomenon is further explored in Sect. 5.1.
The width of the continuum window was set empirically to be wide enough to secure high signal-to-noise ratio fluxes (S/N $\ge 3$)\footnote{More than 99\% of the continuum measurements for the variable targets have a S/N of 3 or higher.} while at the same time minimizing the contribution from contaminants.

\begin{figure}[tp]
\centering
\includegraphics[width=0.45\textwidth, page=1]{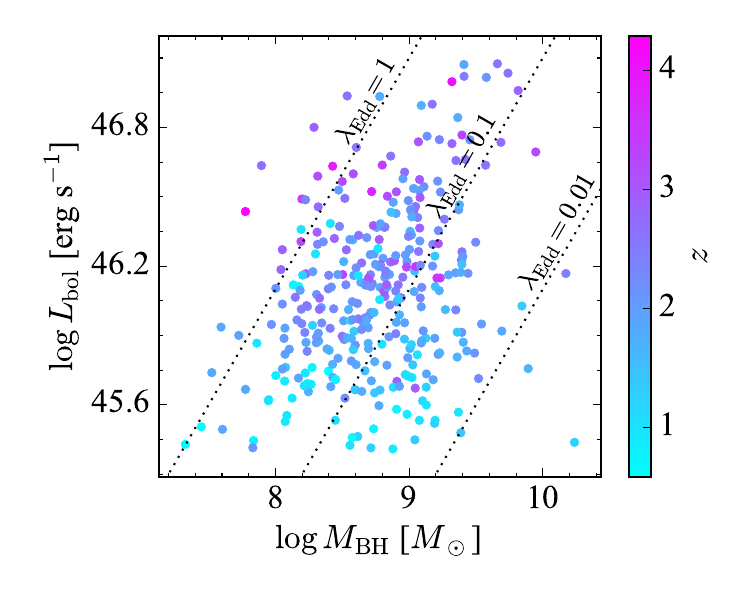}
\caption{Distribution of the BH mass and bolometric luminosity of the final sample with available BH mass measurements. The points are color-coded by redshift. The dotted lines mark Eddington ratios of $\lambda_{\rm Edd} = 0.01$, $0.1$, and $1$.
}
\end{figure}

\subsection{Sample selection}
Before SF analysis, radio-loud sources ($\sim2$\% of the initial sample), classified as those with flux density ratios between 6 cm and 2500 ${\rm \AA}$ (adopted from \citealt{shen_2019}) higher than 10 \citep{kellermann_1989}, were eliminated because of their distinctive variability characteristics (e.g., \citealt{wagner_1995}). We also discarded light curves that lack evidence of variability based on the variability probability criterion $P_{\rm var} < 0.95$, where $P_{\rm var}$, calculated from the $\chi^2$ distribution \citep{mclaughlin_1996, sanchez_2017}, is the probability that the target is intrinsically variable. We computed $P_{\rm var}$   for each continuum window using the $\sim7$ yr light curves;  $P_{\rm var}$    is commonly employed to confirm the intrinsic variability of the continuum flux through the $\chi^2$ distribution. We note that S/N2, defined as $\sqrt{\chi^2 - {\rm DOF}}$, where DOF is the number of degrees of freedom of the light curve, is typically used to quantify the variability of the emission flux (\citealt{grier_2019}). However, we chose not to use S/N2 in this study, as the criteria for the intrinsic variability were determined through visual inspection. To compute ensemble SFs (see Sect.~3), we divided the sample based on the bolometric luminosity,starting from \llbol\ = 45.4 in steps of 0.25. Objects with luminosities lower than \llbol\ $\leq45.4$ were removed because they can be substantially contaminated by the host galaxy ($f_{\rm host} \leq 0.1$ at 3450 \AA; \citealt{shen_2015b}), even though we only use rest-frame ultraviolet wavelengths. The final sample consists of 447 AGNs that have light curves in at least one continuum window among 1367, 1479, 1746, 2230, 3050, 3500, and 3800 ${\rm \AA}$. In total, 50, 85, 205, 322, 310, 249, and 197 light curves are available, respectively, for the continuum windows at 1367, 1479, 1746, 2230, 3050, 3500, and 3800 ${\rm \AA}$.

\subsection{AGN properties}
To calculate the bolometric luminosity of the AGN, we adopted the bolometric correction factors \lbol\ $= 3.81\, L_{\rm 1350\, \AA}$ and \lbol\ $= 5.15\, L_{\rm 3000\, \AA}$ \citep{richards_2006,shen_2011}. We used the median values of the monochromatic luminosities at 1350 ${\rm\AA}$ and 3000 ${\rm\AA}$ measured from 2014 to 2020 as part of the SDSS-RM project \citep{shen_2024}, giving priority to the former over the latter since the bolometric correction factor for $L_{\rm 3000 \AA}$ may not be constant and could vary with the AGN luminosity \cite[e.g.][]{trakhtenbrot_2012}. BH masses were estimated based on the RM method using \hal, \hb, \mgii\ $\lambda 2800$, or \civ\ $\lambda 1549$. For AGNs without successful measurements of time lags from RM, we adopted BH masses based on the single-epoch spectrum of \hb\ or \civ. We chose not to use BH mass estimates based on the single-epoch spectrum with \mgii\ because it appears to yield systematically lower values than that from \mgii\ RM. Our final sample of 447 quasars spans redshifts $z = 0.58 - 4.29$, AGN bolometric luminosities $L_{\rm bol} = 10^{45.4}-10^{47.1}$\,erg\,s$^{-1}$, BH masses $M_{\rm BH} = 10^{7.3}-10^{10.2}\,M_\odot$, and Eddington ratios \redd\ $\equiv L_{\rm bol}/L_{\rm Edd} = 10^{-2.9}-10^{0.6}$ (Fig.~2), with Eddington luminosity \ledd\ = $1.26\times10^{38}\,M_{\rm BH}/M_\odot\,{\rm erg\,s}^{-1}$.

\section{Ensemble structure function}
Following the definition from \cite{press_1992}, the SF of an individual object was calculated as
\begin{eqnarray}
    {\rm SF}^2(\Delta t) = \frac{1}{N_{\Delta t, {\rm pair}}} \sum_{i=1}^{N_{\Delta t, {\rm pair}}}
    (m(t) - m(t+\Delta t))^2 \nonumber \\
    - \sigma_{\rm final}^2(t) - \sigma_{\rm final}^2(t+\Delta t), 
\end{eqnarray}
\noindent
where $N_{\Delta t, {\rm pair}}$ is the number of observed pairs with the time lag ($\Delta t$), $m$ is the observed magnitude, and $\sigma_{\rm final}$ is the final uncertainty of magnitude in the single-epoch spectrum, as described in Sect.~2.1. We adopted this SF definition because incomplete noise terms often lead to a flatter SF than the real SF \citep{kozlowski_2016b}. Because the SF can be highly sensitive to outliers, we applied $3\,\sigma$ clipping to the median with half of the difference between the 16th and 84th percentile of magnitude measurements in each light curve.

To examine the average SF as a function of luminosity and wavelength, we used an ensemble SF instead of the SF from individual objects. The ensemble SF was calculated by averaging SF$^2$s of sources in each luminosity and wavelength bin, considering that the SFs of individual sources are often associated with systematically large uncertainties and have negative values \cite[e.g.,][]{son_2023b, kim_2024}. To avoid potential bias due to the small sample size, the ensemble SF was computed only if there were at least ten objects in a group. The number of objects used to estimate the ensemble SF and AGN properties for each group is provided in Appendix A. The derived ensemble SFs in different luminosity and wavelength bins are shown in Fig.~3.

To evaluate potential biases in estimating the ensemble SFs caused by the seasonal gaps and the limited baseline in the light curves, we performed the simulations using regularly sampled light curves with a cadence of one day, assuming the damped random walk model. Subsequently, we generated mock light curves, incorporating the actual cadence and redshift distributions of our sample. This experiment demonstrates that the ensemble SFs for $\Delta t  \lessapprox 200$ days can be reproduced accurately using the simulated light curves \cite[see also][]{emmanoulopoulos_2010,kozlowski_2017}.

\section{Results}

To understand the physical origin of the variability, we examined how the characteristic timescale depends on the physical properties of AGNs. Previous studies demonstrated that $\tau$ might correspond to the thermal timescale \citep{kelly_2009,kelly_2013,caplar_2017,tang_2023,arevalo_2024}. In the standard disk model \citep{shakura_1973, frank_2002}, the thermal timescale correlates with the orbital timescale ($t_{\rm orb}$):

\begin{equation} \label{eq:thr}
t_{\rm th} \propto t_{\rm orb}\,\alpha^{-1} \propto R^{3/2}_\lambda\,M^{-1/2}_{\rm BH}\,\alpha^{-1},
\end{equation}

\noindent
where $R_\lambda$ is the characteristic disk radius responsible for continuum emission at the specific rest-frame wavelength, and $\alpha$ is the viscosity parameter, assumed constant in the standard disk model. Since the disk radiates blackbody emission, its effective temperature ($T$) at a given disk radius is given by 

\begin{equation}
T(R) \propto (M_{\rm BH}\,\dot{M}\,R^{-3})^{1/4}, 
\end{equation}

\noindent
where $\dot{M}$ is the mass accretion rate and $R$ is the distance from the center of the disk. Using Wien's displacement law, $\lambda_{\rm max} \propto T^{-1}$, 

\begin{equation}
R_{\lambda} \propto M^{1/3}_{\rm BH} \dot{M}^{1/3} \lambda^{4/3}_{\rm RF}.
\end{equation}

\noindent
From Eqs. (2) and (4) with an assumption of constant $\alpha$ and $L \propto \dot{M}$, the thermal timescale is correlated with the luminosity and rest-frame wavelength ($\lambda^2_{\rm RF}$) as 

\begin{equation} \label{eq:tth}
t_{\rm th} \propto L^{1/2} \lambda^2_{\rm RF}.
\end{equation}

\noindent
To test whether this holds true for our dataset, we examined the dependence of $\tau$ on the AGN luminosity and wavelength in the form 

\begin{equation} \label{eq:fit1}
\tau \propto L^{a}_{\rm bol}\,\lambda^{b}_{\rm RF}.
\end{equation}

The light curves are insufficiently long to constrain $\tau$ directly. Instead, we adopted an indirect method to investigate the dependence of $\tau$ on the AGN luminosity and wavelength, assuming that all the SFs follow a single universal function if $\Delta t$ is converted to units of $\tau$ \citep[e.g.,][]{tang_2023, arevalo_2024, petrecca_2024} as

\begin{equation} \label{eq:sf}
{\rm SF} = \delta (\Delta t/\tau)^\gamma,
\end{equation}

\noindent
where $\gamma$ is the power-law slope of the SF and $\delta$ is the SF at $\Delta t=\tau$. Since $\tau$ is derived in arbitrary units, the value of $\delta$  has no physical meaning in this study. Fitting the ensemble SFs of each luminosity group at different wavelengths, we can investigate the dependence of $\tau$ on luminosity and wavelength. To achieve this, we first computed the ensemble SFs within luminosity bins for each continuum window and then fit them using the Python package \texttt{lmfit} to determine $a$ and $b$.

We derive $a = 0.50\pm0.03$ and $b = 1.42\pm0.09$ (Fig.~3b). While the best-fit value of $a$ is in excellent agreement with the predicted value of 0.5 from the standard disk model, the value of $b$ is substantially smaller than the predicted value of 2, by more than $6\,\sigma$. This reveals that the temperature (color) radial profile of the accretion disk deviates from the standard disk model. 

\begin{figure*}[htp!]
\centering
\includegraphics[width=0.9\textwidth]{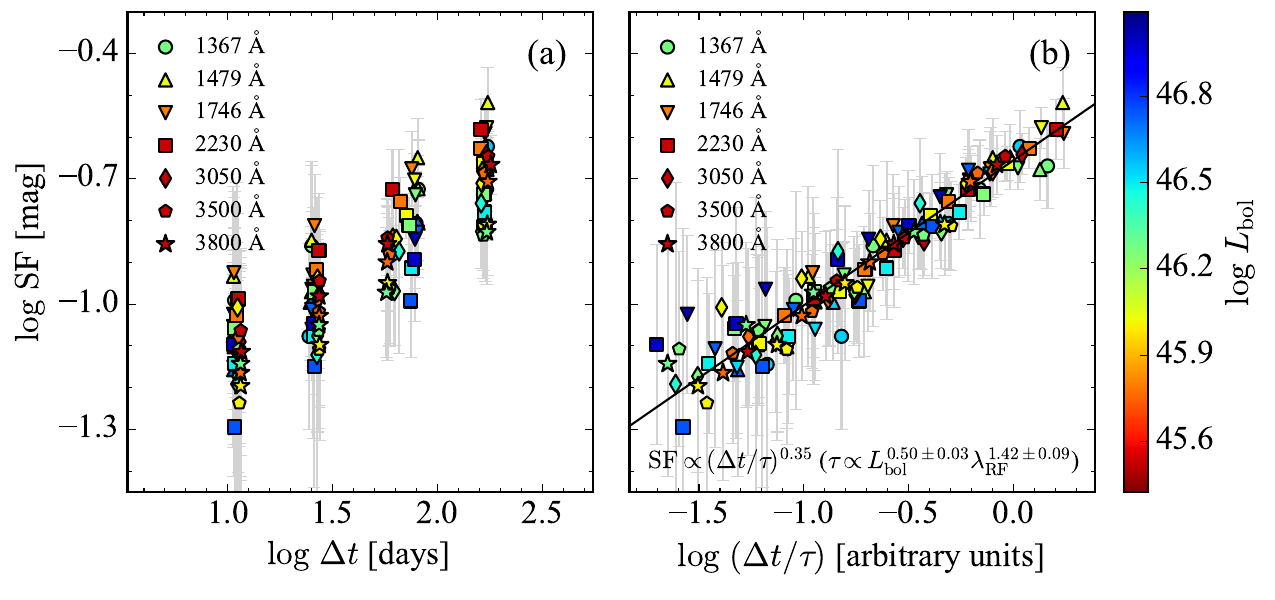}
\caption{Ensemble SF as a function of $\Delta t$ (a) and $\Delta t/\tau$  (b) at different wavelengths for each group, color-coded by the average bolometric luminosity. Different symbols indicate the different wavelengths used for continuum extraction. Panel (b) assumes that the SFs all follow a single universal function if $\Delta t$ is converted to units of $\tau \propto L_{\rm bol}^a \lambda_{\rm RF}^b$. Note that the values of $\Delta t/\tau$ on the x-axis are only meaningful in relative terms. The solid black line represents the best-fit universal SF.
}
\end{figure*}

\section{Discussion}
\subsection{Possible origins of the distinctive color profile}

The standard accretion disk is known to maintain its structure when the AGN is accreting at intermediate Eddington ratios of \redd\ $\approx 0.01-0.3$. However, at least 100 objects in our final sample have \redd\ $\gtrsim 0.3$, and a handful of sources (four) have \redd\ $<0.01$ (Fig. 2). We note that this is the minimum number because the Eddington ratio can only be calculated for the final sample with available BH mass measurements (i.e., a BH mass based on the RM technique or broad \civ\ and \hb\ emissions from a single-epoch spectrum). When \redd\ $\gtrsim 0.3$, the cooling in the inner part of the disk is dominated by advection, which results in an optically and geometrically thick disk (a ``slim disk'': \citealt{begelman_1978,abramowicz_1988,wang_2014}). At the opposite extreme, when the accretion rates drop to low enough values and \redd\ $\lesssim0.01$, the accretion disk becomes optically thin and geometrically thick, transforming into a radiatively inefficient accretion flow (RIAF; \citealt{yuan_2014, ho_2008}). The majority of central BHs in the local Universe reside in this state \citep{ho_2009,she_2017}. We tested whether the distinctive color profile with $\tau \propto \lambda_{\rm RF}^{1.42}$ is due to the inclusion of objects with non-standard disks.

Here, we only focused on the impact of slim disks, as they account for nearly one-quarter of the final sample; highly sub-Eddington systems in the RIAF state account for less than 1\% of our sources. Considering that a slim disk has a distinctive geometry, temperature profile, and kinematics compared to the standard disk, its characteristic timescale can differ from that of a standard disk \citep{beloborodov_1998,wang_1999,watarai_2000}. However, the state change mostly occurs inside the photon-trapping radius of the slim disk, where the accretion disk becomes vertically extended due to the radiation pressure (\citealt{abramowicz_1988,wang_1999,sadowski_2011}). To test if this phenomenon is directly related to our finding, we first computed the photon-trapping radius and compared it with the effective emitting radius corresponding to the continuum wavelengths used in this study. The self-similar solution of \cite{wang_1999} for a slim disk expresses the photon-trapping radius as

\begin{equation} \label{eq:Rtr}
R_{\rm tr} = 0.9\,\dot{m}\,R_g,
\end{equation}

\noindent
where $\dot{m} = \dot{M}/\dot{M}_{\rm Edd}$ is the dimensionless accretion rate and $R_g= 2GM/c^2$ is the Schwarzschild radius. In the same model, the effective temperature at a given radius ($R$) is given by

\begin{equation}
T = 2.1 \times 10^7\,(M_{\rm BH}/M_\odot)^{-1/4}(R/R_g)^{-1/2}.
\end{equation}

\noindent
Assuming that the slim disk radiates blackbody emission similar to a standard disk, using Wien's displacement law ($\lambda_{\rm max}\,T = 2.898\times10^{-3}\,{\rm m\,K}$), the photon-trapping wavelength, where the radiation from the photon-trapping radius peaks, is calculated as \citep{wang_1999,cackett_2020}

\begin{equation} \label{eq:lambdatr}
\lambda_{\rm tr} = 1.3\,(M_{\rm BH}/M_\odot)^{1/4}\,\dot{m}^{1/2}\,{\rm \AA}.
\end{equation}

\noindent
We note that $\dot{m}$ cannot be directly measured for our sample because the BH spin and the viscous parameter are unknown. Therefore, we conservatively assumed that the BH spin is 0 and the viscous parameter is 0.01, whereupon $\dot{m}$ is highest and hence the photon-trapping radius is largest at a given Eddington ratio. With these assumptions, we estimated $\dot{m}$ visually from Fig.~4.11 of \cite{sadowski_2011}; the uncertainty of this visual estimate is $\sim 3\%$ ($0.01$ dex). With the estimate of $\dot{m}$ in hand, Eq.~\eqref{eq:lambdatr} can then be used to deduce that the photon-trapping wavelength ($\lambda_{\rm tr}$) for the majority of the sample ranges from $\sim300$ \AA\ to $\sim1350$ \AA, which is significantly shorter than the continuum wavelengths ($1367-3800 {\rm \AA}$) used in this study\footnote{The only exception is a single object for which $\lambda_{\rm tr}=1482\,{\rm \AA}$ is larger than 1367 and 1479 ${\rm \AA}$, and omitting this target hardly changes the fit of Eq. (6).}. The continuum fluxes analyzed here mostly originate from emitting regions outside the photon-trapping radius for highly accreting AGNs. It is also noteworthy that, based on both theoretical \citep{watarai_2000} and observational considerations \citep{li_2024}, the effective temperature profile of a slim disk outside of its photon-trapping radius is not markedly different from that of a standard thin disk. We conclude that the discrepancy of the $\tau - \lambda_{\rm RF}^b$ between this study ($b=1.43\pm0.09$) and the prediction from the standard disk model ($b=2$) is unlikely to originate from the transition to the slim disk. 

To test the impact of RIAFs, we additionally rejected the four AGNs with RIAFs. The fit to Eq. (6) remains almost the same ($a=0.49\pm0.03$, $b=1.43\pm0.09$). As a final sanity check, we applied the fit to the ensemble SFs for objects with $0.01 < \lambda_{\rm Edd} \leq 0.3$ that strictly speaking conform to the expectations of a standard disk, and still the results remain the same ($a=0.46\pm0.07$, $b=1.32\pm0.15$; Table~1). Neither slim disks nor RIAFs have a significant impact on our results.

Another factor that might contribute to the departure from the expected color profile is contamination from the Balmer pseudo-continuum originating from the BLR, which contributes $\sim10\%$ on average to the flux density at 3000 ${\rm\AA}$ \citep{kovacevic_2017}. This pseudo-continuum has also been regarded as a culprit for the disk size problem identified from continuum RM studies (e.g., \citealt{fausnaugh_2016,mchardy_2018}). To evaluate the possible effect of the Balmer continuum on our results, we refit Eq. (6) by excluding wavelengths longer than 3000 ${\rm \AA}$, the spectral range most severely contaminated by the Balmer continuum \cite[e.g.,][]{cackett_2018}. The fitting results hardly change ($a=0.50\pm0.04$, $b=1.45\pm0.18$) within a $1\,\sigma$ level. Therefore, the Balmer continuum does not seem to be the origin of the color profile discrepancy either.

\begin{table}
\caption{Best-fit parameters for $\tau \propto L^a\,\lambda^b_{\rm RF}$.
\label{tab:table2}}
\centering
\begin{tabular}{ccc}
\hline \hline
Subsample &
$a$ &
$b$ \\
(1) &
(2) & 
(3) \\
\hline
All & $0.50\pm0.03$ & $1.42\pm0.09$ \\ 
$0.01 < \lambda_{\rm Edd} \leq 0.3$ & $0.46\pm0.07$ & $1.32\pm0.15$ \\ 
All with $\lambda_{\rm RF} < 3000\,{\rm \AA}$ & $0.50\pm0.04$ & $1.45\pm0.18$ \\ 
\hline
\end{tabular}
\end{table}

\subsection{Comparison with observational studies}
Various studies have investigated the color profiles in luminous AGNs based on light curves from broadband photometric data. For example, using the broadband data from the first year of the SDSS-RM project, \citet{homayouni_2019} found that the color profile ($R\propto \lambda_{\rm RF}^p$) is well fit with $p=1.28_{-0.39}^{+0.41}$, equivalent to $b=3p/2=1.92_{-0.585}^{+0.615}$. Their estimation is consistent with both the standard disk prediction and our best fit, considering their considerably larger uncertainties. \cite{tang_2023}, employing the same method as ours, showed that the characteristic timescale is proportional to $L^{0.539\pm0.004}\lambda^{2.418\pm0.023}$; their values of $a$ and $b$ are formally steeper than those predicted by the standard disk. It is unclear why their estimate of $b$ is significantly larger than that of our study. We speculate that the discrepancy may arise from a potential bias introduced by the broadband photometry data. In addition, \cite{tang_2023} used a longer $\Delta t$ (up to 350 days) than we did (up to 179 days).  
We imposed this $\Delta t$ limit because of the restricted baseline of our dataset (e.g., \citealt{emmanoulopoulos_2010}), which is more pronounced at shorter wavelengths (1367 and 1479 \AA) because of cosmological time dilation. However, two studies based on photometric data offer counter-examples. Analyzing the characteristic timescales of thousands of individual quasars using $\sim10$ years of broadband light curves from the SDSS Stripe 82 region \citep{jiang_2014}, \citet{macleod_2010} found $\tau \propto \lambda_{\rm RF}^{0.17\pm0.02}$. Similarly, \citet[see the erratum in \citealt{stone_2023}]{stone_2022} studied $\sim20$ years of broadband light curves for 190 SDSS Stripe 82 quasars and found $\tau \propto \lambda_{\rm RF}^{0.35\pm0.14}$ or $\lambda_{\rm RF}^{0.43\pm0.01}$ depending on the sample selection. These works produced significantly shallower slopes than our study. The situation for continuum RM based on spectroscopic data is no less confusing. While the work of \cite{fausnaugh_2016} and \cite{starkey_2017} yielded $p=0.99\pm0.14$ or $b=1.49\pm0.21$, which does agree well with our best-fit results, most are at variance, with a preponderance of studies supporting $p \approx 4/3$ \citep{cackett_2018,collier_1998,homayouni_2019,guo_2022}.

Microlensing analysis, too, has produced a wide diversity of power-law indices, ranging from $p\approx 4/3$ (\citealt{eigenbrod_2008,poindexter_2008,rivera_2023}) to $p<4/3$ (\citealt{floyd_2009,blackburne_2011,jimenez_2014,munoz_2016}) and $p>4/3$ (\citealt{bate_2018,cornachione_2020}). It is noteworthy that microlensing studies based on spectroscopic data and/or narrowband photometric data have a tendency to favor smaller power-law index values  ($p\approx 0.7-1.3$ or $b\approx1.05-1.95$; \citealt{eigenbrod_2008,mediavilla_2011,mosquera_2011,motta_2012,jimenez_2014,munoz_2016}) compared to the prediction from the standard disk model; this qualitatively aligns with our study.

Overall, the above examples underscore that estimates of radial temperature profiles of accretion disks can be sensitive to the adopted methodology and sample selection. They also highlight the merits of using spectroscopic data, as adopted in this work, instead of the more expedient broadband photometry.

\subsection{Application to the accretion disk model}
Assuming that AGNs obey a universal SF, we find that the characteristic timescale ($\tau$) is proportional to $L^{0.5}$ and $\lambda_{\rm RF}^{1.42}$. While the luminosity dependence is consistent with the prediction from standard disk theory, the wavelength scaling reveals that the radial temperature profile is considerably steeper than that of a standard disk. After careful consideration of factors that might affect the analysis, we conclude that this discrepancy in the temperature profile is likely genuine.

There is presently no obvious resolution to this dilemma. From a theoretical point of view, a steep temperature (and hence a shallow color) profile can be partly explained by the magnetorotational instability model attributed to additional dissipation originating from torque at the inner edge of the accretion disk. For example, the magnetorotational instability model of \citet{agol_2000} predicts a power-law index of $p=8/7=1.14$ ($b=12/7=1.71$), which is marginally smaller than $p=4/3$ ($b=2$), although not as small as our observed value of $b = 1.42$. {The steep temperature profile can also be produced by the continuum emission at a specific wavelength, which is contaminated by emission from different radii \citep{zhou_2024}. Considering this contamination, \citet{zhou_2024} demonstrated that $a$ and $b$ can be 0.65 and 1.19, respectively, from the simulated mock light curves based on the corona-heated accretion-disk reprocessing (CHAR) model \citep{sun_2020}. While their $a$ is marginally higher than our result, their predicted value of $b$ broadly aligns with our observation within $3\sigma$.}

On the other hand, the plethora of scenarios that have been proposed to solve the disk size problem either do not focus on the temperature profile or predict temperature profiles that contradict our results \cite[e.g.,][]{lawrence_2012,gardner_2017,kammoun_2019,li_2019,sun_2019,sun_2020,starkey_2023}. For instance, \citet{lawrence_2012} argued that reprocessed continuum emission, presumably due to cold, dense clouds located at $\sim 30\,R_S$, may be crucial to explain the disk size problem. Contrary to our results, however, a steeper temperature profile is expected from this model. In a similar manner, the disk wind model favors not only a steeper temperature profile but also a higher power-law index for the AGN luminosity (i.e., $a>0.5$; \citealp{li_2019,sun_2019}). The temperature profile of the rimmed accretion disk model of \citet{starkey_2023} does not depart substantially (i.e., $T\propto R^{-3/4}$) from that of the standard disk, except for the outer part of the disk.

\section{Conclusion}
We constructed ultraviolet ensemble SFs of luminous AGNs grouped by luminosity and continuum wavelength using spectrophotometric data obtained through the SDSS-RM project. Assuming that all AGNs share a single universal SF after $\Delta t$ is normalized by the characteristic timescale ($\tau$), we fit all the ensemble SFs to investigate how the characteristic timescale depends on luminosity and wavelength in the form $\tau \propto L^a \lambda_{\rm RF}^b$. Our main findings are:

\begin{itemize}

\item The ensemble SFs are well fit by a single universal SF($\Delta t/\tau$) with $a=0.50\pm0.03$ and $b=1.42\pm0.09$.

\item The dependence of $\tau$ on luminosity ($a=0.50\pm0.03$) is in excellent agreement with the predicted value of $a=0.5$ from the standard disk model.

\item The dependence of $\tau$ on the continuum wavelength ($b=1.42\pm0.09$) departs by more than $6\,\sigma$ from the standard disk model ($b=2$), revealing a shallower color (steeper temperature) profile. 

\item Highly accreting AGNs and the Balmer continuum have a negligible effect on measurements of $b$, strongly suggesting that the discrepancy between the observed and predicted value of $b$ is genuine.

\item Interestingly, our estimation of $b$ based on spectroscopic variability is in broad agreement with previous continuum RM and microlensing analyses based on spectroscopic monitoring data. This suggests that spectroscopic data may yield more robust results on the radial temperature or color profiles than photometric data.

\end{itemize}

\begin{acknowledgements}
We are grateful to the anonymous referee for constructive comments and suggestions that greatly improved our manuscript.
LCH was supported by the National Science Foundation of China (11991052, 12233001), the National Key R\&D Program of China (2022YFF0503401), and the China Manned Space Project (CMS-CSST-2021-A04, CMS-CSST-2021-A06). This work was supported by the National Research Foundation of Korea (NRF) grant funded by the Korean government (MSIT) (Nos. 2022R1A4A3031306, 2023R1A2C1006261, and RS-2024-00347548). 

\end{acknowledgements}

\bibliography{torus}

\begin{appendix}
\section{AGN properties of the sample}
The properties of AGNs and the number of objects for each subsample are listed in Table~A.1.

\begin{table}[ht]
\caption{Properties of the subsamples.}
\label{tab:table1}
\centering
\begin{tabular}{ccccccc}
\hline \hline
Wavelength & $G$ & $N$ & $z$ & \llbol & $\log\ (M_{\rm BH}/M_\odot$) & \lredd \\
(1) & (2) & (3) & (4) & (5) & (6) & (7) \\
\hline
1367 & 3 & 15(15) & 2.91 & 46.27 & 8.67 & $-0.50$ \\
     & 4 & 15(15) & 3.22 & 46.54 & 8.61 & $-0.17$ \\
\hline
1479 & 2 & 17(17) & 2.63 & 46.04 & 8.66 & $-0.72$ \\ 
     & 3 & 26(26) & 2.77 & 46.25 & 8.77 & $-0.62$ \\
     & 4 & 19(19) & 2.98 & 46.54 & 8.79 & $-0.35$ \\
     & 5 & 10(10) & 2.83 & 46.72 & 9.19 & $-0.58$ \\
\hline
1746 & 1 & 28(28) & 2.09 & 45.80 & 8.63 & $-0.93$ \\
     & 2 & 55(55) & 2.30 & 46.03 & 8.63 & $-0.70$ \\
     & 3 & 60(60) & 2.42 & 46.25 & 8.82 & $-0.67$ \\
     & 4 & 31(31) & 2.59 & 46.52 & 8.87 & $-0.44$ \\
     & 5 & 13(13) & 2.68 & 46.73 & 9.21 & $-0.59$ \\
     & 6 & 10(9) & 2.58 & 46.99 & 9.40 & $-0.50$ \\
\hline
2230 & 0 & 24(6) & 1.44 & 45.53 & 8.34 & $-0.94$ \\
     & 1 & 79(53) & 1.69 & 45.79 & 8.68 & $-0.98$ \\
     & 2 & 82(65) & 1.89 & 46.02 & 8.68 & $-0.76$ \\
     & 3 & 74(59) & 2.04 & 46.25 & 8.88 & $-0.73$ \\
     & 4 & 30(28) & 2.21 & 46.49 & 8.95 & $-0.55$ \\
     & 5 & 15(12) & 2.20 & 46.74 & 9.15 & $-0.50$ \\
     & 6 & 10(9) & 2.36 & 46.99 & 9.34 & $-0.45$ \\
\hline
3050 & 0 & 62(28) & 1.17 & 45.52 & 8.62 & $-1.20$ \\
     & 1 & 74(43) & 1.39 & 45.77 & 8.67 & $-1.01$ \\
     & 2 & 52(30) & 1.51 & 46.02 & 8.74 & $-0.82$ \\
     & 3 & 35(18) & 1.47 & 46.25 & 8.78 & $-0.64$ \\
     & 4 & 10(7) & 1.63 & 46.45 & 9.01 & $-0.66$ \\
\hline
3500 & 0 & 48(23) & 1.09 & 45.53 & 8.60 & $-1.17$ \\
     & 1 & 50(27) & 1.24 & 45.76 & 8.59 & $-0.94$ \\
     & 2 & 25(11) & 1.26 & 46.00 & 8.78 & $-0.86$ \\
     & 3 & 19(8) & 1.28 & 46.27 & 8.63 & $-0.48$ \\
\hline
3800 & 0 & 34(21) & 1.02 & 45.53 & 8.52 & $-1.09$ \\
     & 1 & 32(21) & 1.10 & 45.76 & 8.60 & $-0.94$ \\
     & 2 & 15(8) & 1.12 & 46.00 & 8.66 & $-0.73$ \\
     & 3 & 11(7) & 1.14 & 46.29 & 8.53 & $-0.36$ \\
\hline
\end{tabular}
\tablefoot{
Col. (1): Central wavelength of continuum windows in the rest-frame.
Col. (2):  
Group divided by bolometric luminosity:
0 = $45.4<$ \llbol\ $\leq45.65$, 
1 = $45.65<$ \llbol\ $\leq45.9$,
2 = $45.9<$ \llbol\ $\leq46.15$,
3 = $46.15<$ \llbol\ $\leq46.4$,
4 = $46.4<$ \llbol\ $\leq46.65$,
5 = $46.65<$ \llbol\ $\leq46.9$, and
6 = $46.9<$ \llbol\ $\leq47.15$.
Col. (3): Number of objects; in parentheses is the number of objects with BH mass estimates.
Col. (4): Average redshift.
Col. (5): Average bolometric luminosity.
Col. (6): Average BH mass.
Col. (7): Average Eddington ratio.\\
}
\end{table}

\end{appendix}

\end{document}